\documentclass[acmsmall, screen]{acmart}

\AtBeginDocument{%
  \providecommand\BibTeX{{%
    \normalfont B\kern-0.5em{\scshape i\kern-0.25em b}\kern-0.8em\TeX}}}

\usepackage{algorithmic}
\usepackage{graphicx}
\usepackage{textcomp}
\usepackage{multicol}
\usepackage{multirow}
\usepackage{ragged2e}
\usepackage{tabularx}
\usepackage{array}
\usepackage{cancel}
\usepackage{subfig}
\usepackage{tcolorbox}
\usepackage[htt]{hyphenat}
\usepackage{enumitem}
\usepackage{arydshln}
\begin{document}


\title{\changed{People Still Care About Facts: Twitter Users Engage More with Factual Discourse than Misinformation---A Comparison Between COVID and General Narratives on Twitter}}


\author{Mirela Silva}
\email{msilva1@ufl.edu}
\affiliation{%
    \department{Electrical \& Computer Engineering}
    \institution{University of Florida}
    \streetaddress{P.O. Box 116200}
    \city{Gainesville}
    \state{FL}
    \country{USA}
    \postcode{32611}}
    
\author{Fabrício Ceschin}
\email{fjoceschin@inf.ufpr.br}
\affiliation{%
  \institution{Federal University of Paraná}
  \streetaddress{Cel. Francisco Heráclito dos Santos, 100}
  \city{Curitiba}
  \state{Paraná}
  \country{Brazil}
  \postcode{81530-000}}

\author{Prakash Shrestha}
\email{prakash.shrestha@ufl.edu}
\affiliation{%
    \department{Electrical \& Computer Engineering}
    \institution{University of Florida}
    \streetaddress{P.O. Box 116200}
    \city{Gainesville}
    \state{FL}
    \country{USA}
    \postcode{32611}}
    
\author{Christopher Brant}
\email{g8rboy15@ufl.edu}
\affiliation{%
    \department{Electrical \& Computer Engineering}
    \institution{University of Florida}
    \streetaddress{P.O. Box 116200}
    \city{Gainesville}
    \state{FL}
    \country{USA}
    \postcode{32611}
    }

\author{Shlok Gilda}
\email{shlokgilda@ufl.edu}
\affiliation{%
    \department{Computer and Information Science \& Engineering}
    \institution{University of Florida}
    \streetaddress{P.O. Box 116200}
    \city{Gainesville}
    \state{FL}
    \country{USA}
    \postcode{32611}
    }

\author{Juliana Fernandes}
\email{juliana@jou.ufl.edu}
\affiliation{%
    \department{College of Journalism and Communications}
    \institution{University of Florida}
    \streetaddress{P.O. Box 118400}
    \city{Gainesville}
    \state{FL}
    \country{USA}
    \postcode{32611}}
    
\author{Catia S. Silva}
\email{catiaspsilva@ece.ufl.edu}
\affiliation{%
    \department{Electrical \& Computer Engineering}
    \institution{University of Florida}
    \streetaddress{P.O. Box 116200}
    \city{Gainesville}
    \state{FL}
    \country{USA}
    \postcode{32611}}

\author{André Grégio}
\email{gregio@inf.ufpr.br}
\affiliation{%
  \institution{Federal University of Paraná}
  \streetaddress{Cel. Francisco Heráclito dos Santos, 100}
  \city{Curitiba}
  \state{Paraná}
  \country{Brazil}
  \postcode{81530-000}}
    
\author{Daniela Oliveira}
\email{daniela@ece.ufl.edu}
\affiliation{%
    \department{Electrical \& Computer Engineering}
    \institution{University of Florida}
    \streetaddress{P.O. Box 116200}
    \city{Gainesville}
    \state{FL}
    \country{USA}
    \postcode{32611}}
    
\author{Luiz Giovanini}
\email{lfrancogiovanini@ufl.edu}
\affiliation{%
    \department{Electrical \& Computer Engineering}
    \institution{University of Florida}
    \streetaddress{P.O. Box 116200}
    \city{Gainesville}
    \state{FL}
    \country{USA}
    \postcode{32611}}

\renewcommand{\shortauthors}{Silva et al.}

\thanks{This material is based upon work supported by the National Science Foundation under grants 2028734 and 1662976, Google Security and Privacy Research Award and University of Florida SeedFund OR-DRPD-ROF2020. \changed{This material is based upon work supported by (while serving at) the National Science Foundation}.}

\definecolor{vividblue}{HTML}{0C7BDC}

\newcommand{\changed}[1]{\textcolor{black}{#1}}

\newcommand{\aka}{a.k.a.\ }
\newcommand{\etal}{\mbox{et al.\ }}
\newcommand{\ie}{\mbox{i.e.,\ }}
\newcommand{\eg}{\mbox{e.g.,\ }}
\newcommand{\etals}{\mbox{et al.'s\ }}
\newcommand{\nfeatures}{\changed{126 }}

\newcommand{\commentario}[1]{\textcolor{red}{\textbf{#1}}}
\newcommand{\ToDo}[1]{
    \textcolor{red}{[\textit{ToDo}: \textbf{#1}]}
    }
\newcommand{\XXX}{\textcolor{red}{XXX}\ }

\newcommand{\needsupdate}{\textcolor{red}{\textbf{NEEDS UPDATE!}}\ }


\begin{abstract}
    \changed{Misinformation} entails the dissemination of falsehoods \changed{that leads to the slow fracturing of  society via decreased trust in democratic processes, institutions, and science.} 
The public has grown aware of the role of social media \changed{as a superspreader of untrustworthy information}, where even pandemics have not been immune. 
In this paper, we 
\changed{focus on COVID-19 misinformation and examine a subset of 2.1M tweets to understand misinformation as a function of engagement, tweet content (COVID-19- vs. non-COVID-19-related), and veracity (misleading or factual). 
Using correlation analysis, we show the most relevant feature subsets among over \nfeatures features that most heavily correlate with misinformation or facts. We found that (i) factual tweets, regardless of whether COVID-related, were more engaging than misinformation tweets; and (ii) features that most heavily correlated with engagement varied depending on the veracity and content of the tweet.
}
\end{abstract}

\ccsdesc[500]{Security and privacy~Human and societal aspects of security and privacy}
\ccsdesc[500]{Security and privacy~Social network security and privacy}

\keywords{Engagement, misinformation, social media, machine learning}

\maketitle

\section{Introduction}\label{sec:intro}

Disinformation refers to false or deceptive content distributed via any communication medium (e.g., word-of-month, print, Internet, radio, broadcast) by an adversary who aims to hurt a target (usually a country, political party, or a community) via the spread of propaganda and promotion of societal division, thus casting doubt in democratic processes, government institutions, and on science. Over the past few years, our society has grown wearily aware of the highly polarized schism that has developed beyond the context of mere political discourse. The perceived extremities of our thoughts and opinions are now intimately meshing with falsehoods and outright lies, calling into question the integrity of our government agencies', political representatives', and our own individual handling of public health crises, such as the COVID-19 pandemic~\cite{Tagliabue2020-nc}. 

The pervasive spread of disinformation we have been witnessing today is not a new phenomenon. The Active Measures disinformation campaigns employed during the Cold War \cite{Bittman1972-zz, rid2020active} bear a disturbing resemblance to what we are witnessing today. We are immersed in an environment of highly polarized, scoop-hungry media, with some outlets spreading demonstrably false information. Twitter and Facebook have become the 21st century version of Cold War balloons spreading disinformation beyond the Iron Curtain.  The Active Measures disinformation campaigns also leveraged science as an indirect target to harm the reputation of the US and its Western allies. One of the first measures, leveraging forged documents and false testimonies, alleged that the US army in Korea was using bacteriological weapons against the enemy front~\cite{Bittman1972-zz}.
\changed{As another example,} the 1980s brought a series of science-related disinformation campaigns, such as the CIA's factory for weaponized mosquitoes~\cite{rid2020active}, or the ``AIDS made in the USA''  narrative, spread by the KGB~\cite{Bittman1972-zz} and which appeared more than forty times in the international press~\cite{United_States_Department_of_State1987-qc}.

The 21st century disinformation campaigns have largely leveraged the affordances of social media platforms~\cite{Mueller2019-li, arif18}, which allow narratives to spread quickly, complicating attribution. Following in the steps of history, during the markedly turbulent times innate to a pandemic, it is sadly unsurprising that, in June of 2020, the Associated Press reported that Russian intelligence services are behind the spread of disinformation about the \changed{novel} coronavirus~\cite{Tucker2020-kt}. 

\changed{\textit{Mis}information, however, is closely related to disinformation, and differs only in the lack of purposeful intent to do harm, often coupled with the raw ignorance of the individual spreading such misleading facts.}
COVID-19-related misinformation is therefore \changed{largely} coming from domestic sources, as we have seen politicians, pundits, and personalities pushing \changed{misleading} narratives~\cite{Bell2020-xj} that may prevent society from controlling the spread of the coronavirus, potentially increasing the number of deaths. 
With \changed{the advent of the COVID vaccines}, \changed{misinformation has} unequivocally been used to discredit its effectiveness, preventing efficient immunization and fueling further hyperpartisanship. 

\changed{We posit that engagement is a pivotal dimension towards the dissemination of falsehoods; Avram et al.~\cite{avram2020exposure}, for example, illustrated that higher levels of social engagement tend to result in less fact checking and verification, especially for content with lower credibility.}
In this paper, we aim to investigate the relationship between misinformation and user engagement in COVID-19-related tweets. We opt to use the term \textit{misinformation} instead of disinformation because the latter implies purposeful malice. When false information is spread unintentionally---for example, when a user is successfully deceived by a piece of disinformation, and naively retweets the falsehood---this spread of deceptive content is instead referred to as \textit{misinformation}. For the remainder of this paper, we will use the term misinformation to refer to tweets spreading deceptive content, even though some misinformation tweets might have been created with malice. 
Using \changed{a curated dataset of 2.1M tweets} with the labels of \texttt{fact} and \texttt{misinformation} \changed{for COVID-related tweets and tweets pertaining to other, ``general topics''}, we sought to answer the following research questions:

\begin{itemize}[leftmargin=*]
    \item\textbf{RQ1:}  Are \changed{COVID-19} misinformation tweets more engaging than \changed{COVID-19} factual tweets?
    \item\textbf{RQ2:} Are \changed{general topic} misinformation tweets more engaging than \changed{general topics} factual tweets?
    \item\textbf{RQ3:} Which features are most \changed{correlated with engagement in COVID-19 vs. general topics misinformation tweets?}
    \item\textbf{RQ4:} Which features are most \changed{correlated with engagement in COVID-19 vs. general topics factual tweets?}
\end{itemize}

To do so, we measured engagement as the summation of \# likes and \# of retweets \changed{(i.e., \textit{combined engagement}}. After completing textual preprocessing on the tweets themselves, we submitted \changed{$10$ stratified random samples of our dataset} to several \changed{statistical and correlational analyses to study the relationship between combined engagement and \nfeatures features}. 
We found that:
(i) factual tweets were more engaging than misinformation tweets, \changed{regardless of the topic (i.e., COVID or general) of the tweets}; 
(ii) \changed{features strongly correlated with engagement varied depending on the veracity and topic of the tweets; yet
(iii) features relating to the syntactical content of the tweet (e.g., use of informal speech, punctuation) were strongly correlated with general and COVID-related factual tweets, as well as misinformation COVID-related tweets while
(iv) user metadata (e.g., whether the user account was verified) strongly correlated with general topic misinformation, but not COVID-19 misinformation; and
(v) semantic features (e.g., sentiment, writing with clout/confidence) were strongly correlated with factual COVID-related tweets, but not misinformation.
All of these findings show that understanding and addressing misinformation should potentially take a more targeted direction depending on the issue, instead a one-size-fits-all approach.
}

Research on COVID-19 \changed{misinformation} is in its infancy as most relevant research papers on COVID-related misinformation are \changed{still} pre-prints focused on the detection, prevalence and sentiments of COVID-19 misinformation in social media~\cite{singh2020first, schild2020go, cinelli2020covid, sharma2020covid, memon2020characterizing, yang2020prevalence, huang2020disinformation}), \changed{with many works focusing on} Twitter users' perception of the pandemic~\cite{roozenbeek2020susceptibility, swami2020analytic}. To the best of our knowledge, \changed{no prior work has studied users' engagement as it related to factual and misinformation tweets, relative to COVID- and general-related topics.}
This paper thus makes the following contributions:

\begin{enumerate}[leftmargin=*]
    \item We analyze Twitter discourse on COVID-19 \changed{and non-COVID-19 topics} to discover whether misinformation tweets are more engaging than factual tweets.
    \item We identify discriminating characteristics of a tweet and its author that can distinguish factual and misinformation tweets \changed{based on} tweet engagement.
    \item We will \changed{publicly} release our curated dataset\footnote{In order to comply with Twitter's Terms of Service (\url{https://developer.twitter.com/en/developer-terms/agreement-and-policy}), we omitted the tweet's raw text, as well as any features that could potentially reveal the users' identity.} containing a total of \nfeatures features and labels obtained from several analyses (factual/misinformation-, sociolinguistic-, morality-, and sentiment--focused) for the \changed{$\sim$2.1M} COVID-19 and general topics tweets originally made available \changed{across nine different datasets to foster more research on this topic}.
\end{enumerate}

Though technological advancements enable better disinformation campaigns that can reach more people faster and complicate attribution, better technology also gives society better tools for defense against this threat: faster exposure, detection of \changed{misinformation} seeds, and better reach for awareness campaigns. 
Understanding how automated solutions can distinguish misinformation and what makes for engaging misinformation can aid future defense approaches. Furthermore, insights about features predicting engagement in health-related  topics in short social media texts can be leveraged by public health organizations and officials to convey important public health information to society.

Our paper is organized as follows. 
Section~\ref{sec:related_works} reviews prior works on misinformation and public health, and considers the added value of our work. 
Section~\ref{sec:dataset} discusses our dataset, its curation process, and the preprocessing and feature extraction steps taken. 
Section~\ref{sec:analysis_results} then analyses our cleaned dataset's results via statistical tests. 
Section~\ref{sec:discussion} discusses the take-aways and limitations of our analyses, as well as the future directions for this line of work. 
Section~\ref{sec:conclusion} concludes the paper. 

\section{Related Works}\label{sec:related_works}

\changed{
With the goal of understanding the nuances that correlate engagement to COVID-19 and other topics of misinformation in the Twittersphere, there have been a few unique approaches that have produced intriguing results.
}
In this section, we provide an overview of literature relevant to our work.


Various recent researchers have also explored the presence, prevalence, and sentiment of misinformation on social media of COVID-19 discourse~\cite{sharma2020covid, huang2020disinformation, singh2020first, cinelli2020covid, yang2020prevalence, memon2020characterizing, al2020lies}, user's susceptibility and psychological perceptions on this public health crisis~\cite{roozenbeek2020susceptibility, swami2020analytic}, the predictors of fake news~\cite{apuke2020fake, islam2020misinformation}, and the role of bots~\cite{memon2020characterizing, yang2020prevalence} on spreading COVID-19 misinformation. 
For instance, Sharma et al.~\cite{sharma2020covid} examined Twitter data to identify misinformation tweets leveraging state-of-the-art fact-checking tools (e.g., Media Bias/Fact Check, NewsGuard, and Zimdars) along with topics, sentiments, and emerging trends in the COVID-19 Twitter discourse. 
Singh et al.~\cite{singh2020first} found that misinformation and myths on COVID-19 are discussed at a lower volume than other pandemic-specific themes on Twitter. 
They also concluded that information flow on Twitter shows a spatio-temporal relationship with the infection rates. 
Jiang et al.~\cite{jiang2020political} examined the usage of hashtags in 2.3M tweets in the United States, and observed that the American public frames the pandemic as a core political issue. 
Cinelli et al.~\cite{cinelli2020covid} went beyond Twitter and analyzed data from four other social media platforms: Instagram, YouTube, Reddit, and Gab, finding different volumes of misinformation on each platform.
\changed{
Huang et al.~\cite{huang2020disinformation} investigated $\sim$67.4M COVID mis- and disinformation across seven user social categories (e.g., government official, regular user, etc.), 
finding that
}
news media and government officials tweets are highly engaging, and that most discussion on misinformation originates from the United States.
Unlike this work that explored the kind of users involved in and the location of dissemination of highly engaging tweets, this present paper aims to identify the set of tweet and user characteristics that can predict factual/misinformation tweets and engagement with factual/misinformation tweets.
Although several works have been conducted on COVID-19 misinformation, \changed{almost none} of these studies focused on measuring users' engagement and discriminating features for engagement in COVID-19-related misinformation and information tweets, as proposed in this paper. 

\changed{
One notable exception is the work by Al-Rakhami and Al-Amri~\cite{al2020lies}, which utilized 409K COVID-related tweets collected from January to April 2020, and entrusted 10 human annotators with labeling the tweets are credible (i.e., from a reliable source such as the WHO or a fact-checking platform) or uncredible (i.e., bearing false information). 
The authors then collected a total of 26 user- (e.g., \# followers) and tweet-related (e.g., \# hashtags) features, and used entropy- and correlation-based ranking for selecting the top five features in distinguishing between misinformation and factual information (i.e., whether the user account is verified or not, number of mentions, number of hashtags, number of retweets, and following rate). 
While we curated a feature list containing many of Al-Rakhami and Al-Amri's features, we further expanded the list to contain a total of \nfeatures features that also describe the textual content of the tweet itself.
We also leveraged these features not with the goal of distinguishing mis- from factual information, but instead to understand which features most contribute to engagement. 
Al-Rakhami and Al-Amri also developed their methodology based on the assumption that Twitter users with large followings will likely refrain from spreading misinformation~\cite{Alrubaian2018-kb}.
Today, however, the landscape of misinformation may have shifted; verified Twitter users appear to have shared an all-time high amount of misinformation in 2020~\cite{Cohen2021-nh}, and the Center for Countering Digital Hate identified the ``Disinformation Dozen,'' i.e., 12 \textit{anti-vaxxers} who appear to be responsible for producing 65\% of the shares of vaccine hesitancy misinformation on social media~\cite{Center_for_Countering_Digital_Hate2021-bd}.
Indeed, we found a seemingly positive correlation between the number of followers a user has and engagement with general topics misinformation. 
}

\changed{
Some studies have analyzed engagement metrics in the context of misinformation on social media.
Vosoughi et al.~\cite{vosoughi2018spread} notably conducted an analysis of 126K unique stories tweeted by over 3M people between 2006 and 2017, finding strong correlations that fake news spreads further and faster than verified news. 
Their findings suggest that fake or false news tends to have higher engagement than verified news.
The authors' results contrasts with ours: factual tweets for general and COVID-related tweets were found to be more engaging compared to misinformation. 
However, there are several methodological differences between our works that could explain this discrepancy.
First, Vosoughi et al. measured the diffusion (i.e., spread) of fake news as measured relative to retweet count, whereas we measure engagement as a combination between retweets and favorites. 
We also leveraged a much larger dataset of 2.1M tweets, resulting from the curation of nine unique datasets containing COVID- and non-COVID-related false and factual information.
The authors also looked specifically at fake \textit{news} with verified true/false URLs; although our dataset did contain real and fake news (e.g., Dataset 8 as described in Sec.~\ref{sec:dataset}), we also analyzed tweets and replies made by ``regular'' users that did not contain URLs. 
Lastly, as detailed in Sec.~\ref{sec:dicussion_limitations_future}, we were unable to collect several tweets of our curated datasets using the Twitter API prior to settling on the 2.1M total tweets value.
This could indicate that, in the three years since Vosoughi et al.'s work, Twitter may have improved its ability to, at the very least, cull high engagement misinformation tweets.  
}

\changed{
Aldous et al.~\cite{aldous2019view} quantified engagement in social media across four levels: views, likes, comments/shares, and external or cross-posting to different sites.
Their study collected over 2K random news articles posted on Reddit, 2K random news articles posted elsewhere amongst social media platforms, and afterwards extracted features from each article. 
Subsequently, they used three different ML models to determine which features were pertinent for predicting different engagement levels, finding that language, topic, textual, and sentiment features were all significantly relevant to predicting engagement.
However, their work focuses specifically on the features pertaining to the text content of the social media post, and do not consider multimedia content, external context, or attributes pertaining to the author or users who share the articles. 
In contrast, our work was conducted on a much broader set of features (i.e., sociolinguistic and moral framing of the texts, user and tweet metadata) to reduce this limitation.
}

\changed{Many works have also investigated the relationship between a user's personal characteristics and their susceptibility to misinformation.}
Apuke and Omar~\cite{apuke2020fake} and Islam et al.~\cite{islam2020misinformation} used online surveys to model predictors of fake news dissemination on social media. 
While Apuke and Omar found that altruism, information sharing and seeking, socialization, and pass time predict fake news sharing (altruism being the strongest individual predictor), Islam et al. found that self-promotion, entertainment, and lack of self-regulation predict misinformation sharing (deficient of self-regulation being the strongest predictor). 
Roozenbeek et al.~\cite{roozenbeek2020susceptibility} conducted a large international survey in the UK, USA, Ireland, Spain, and Mexico to investigate susceptibility to misinformation on COVID-19 and its influence on key health-related behaviors.
Their survey showed that the majority of people self-report low susceptibility to misinformation; however, a substantial segment of the public were consistently susceptible to certain misinformation claims. They also found a clear association of belief in misinformation with vaccine hesitancy and reduced self-protective measures. 
Similar association was found in the survey of Swami and Barron~\cite{swami2020analytic}, where they conducted a survey with a nationally representative sample of 530 adult users in the United Kingdom. 
\changed{Though this is a large piece of the puzzle in fighting against the pernicious spread of falsehoods,} contrary to these works that investigated users' personal characteristics that drive the users towards sharing misinformation on social media, the present paper focuses on determining what characteristics of misinformation itself make it most influential and engaging.

\section{COVID Misinformation and Factual Datasets: Preprocessing and Feature Engineering}\label{sec:dataset}
\changed{
We curated data from multiple sources to compose four Twitter datasets used in our analysis for this paper: 
(1) COVID-19 misleading claims,
(2) COVID-19 factual claims, 
(3) misleading claims on general topics, and 
(4) factual claims on general topics. 
We specifically combined different sources of data in each dataset to avoid biasing the results and to improve the generalizability of our findings. 
For example, our datasets of COVID-19 claims include discourse related to the spread of the virus, vaccine, etc. 
The two latter datasets were created with the aim of understanding how user engagement with COVID-19 claims (misleading and factual) differs from engagement with other types of claims (e.g., politics, violence, terrorism). 
This section details our process for building the four aforementioned datasets, along with the steps taken for data preprocessing and feature extraction.
}

\subsection{Dataset Selection}
\changed{
Several Twitter datasets can be found in the literature, with some specifically designed for misinformation analysis. 
These datasets include ground truth labels of \texttt{true/factual} and \texttt{fake/misleading} for tweets, replies, and/or news articles included in the tweets via URLs. 
The ground truth labels are typically assigned manually through human annotators; however, automatic annotation strategies are sometimes employed to reach a larger amount of labeled data. Below, we discuss publicly available Twitter datasets for misinformation analysis on different narratives (including COVID-19) and how we leveraged them to compose the datasets used in our analysis.
}

\subsubsection{COVID-19 Tweets}
\changed{
We found five Twitter datasets potentially relevant for the analysis of COVID-19 misinformation, which we combined to compose our datasets of COVID-19 \texttt{misleading} and \texttt{factual} claims.
}

\medskip\noindent
\changed{\textbf{Dataset 1.} 
This dataset\footnote{\url{https://github.com/Gautamshahi/Misinformation_COVID-19}} was released by Shashi et al.~\cite{shahi2021exploratory} and contains $1,736$ tweets spreading Coronavirus-related news articles that have been fact-checked by over 92 professional fact-checking organizations and mentioned on Snopes and/or Poynter. 
These tweets are dated between January and July 2020, and classified into four categories: 
(i) \textit{false} ($N=1,345$), the tweet mentions an article whose claims are untrue; 
(ii) \textit{partially false} ($N=315$), the mentioned article contains a mix of true and false information; 
(iii) \textit{true} ($N=41$), the tweet mentions factual news; and 
(iv) \textit{other} ($N=35$), the mentioned news cannot be categorized due to lack of evidence about its claims. 
As we are specifically interested in factual and misleading claims, we considered only the tweets from the three first categories; therefore,  \textit{false} and \textit{partially false} tweets were included in our dataset of COVID-19 \texttt{misleading} claims, while the \textit{true} tweets were included in our dataset of COVID-19 \texttt{factual} claims. 
}

\medskip \noindent
\changed{
\textbf{Dataset 2.}
This dataset\footnote{\url{https://datasets.simula.no/wico-graph/}} pertains to COVID-19 conspiracy theories linked to 5G networks. 
To build this dataset, Schroeder et al.~\cite{schroeder2020wgraph} collected tweets containing COVID-related keywords tweeted between January and May, 2020, and then filtered only for those mentioning 5G.
Next, they manually labeled a randomly selected sample of $3,000$ tweets as \textit{5G-corona conspiracy}, \textit{other conspiracy}, or \textit{non-conspiracy}. 
Tweets from the first category made misleading claims about the connection between COVID-19 and 5G networks, such as the unfounded idea that 5G networks caused the pandemic by weakening people's immune systems. 
Tweets labeled as \textit{other conspiracy} contained other types of COVID-19 conspiratory claims, such as ideas related to harmful vaccination and intentional release of the virus. 
Lastly, \textit{non-conspiracy} tweets contained non-COVID-related claims, including the notion that 5G networks are dangerous; tweets mentioning or mocking the existence of conspiracy theories also fall into this third group. 
After that, the authors extracted subgraphs from the three mentioned groups and labeled the tweets accordingly (i.e., automatically labeling). 
In total, Schroeder et al. released $\sim$19K tweets promoting COVID-related 5G conspiracies, $\sim$38.7K tweets promoting other COVID-related conspiracies, and $\sim$157K tweets not promoting conspiracies. 
For our study, we included tweets from the first two groups as part of our COVID-19 \texttt{misleading} claims dataset. We opted to discard tweets not promoting conspiracies because they mixed factual (e.g., existence of Coronavirus-related conspiracies) and misleading  (e.g., 5G networks are harmful) claims.
}

\medskip \noindent
\changed{
\textbf{Dataset 3.} 
The \underline{Co}vid-19 He\underline{a}lthcare M\underline{i}sinformation \underline{D}ataset (CoAID)\footnote{\url{https://github.com/cuilimeng/CoAID}} released by Cui and Lee~\cite{cui2020coaid} includes news articles and social media posts related to COVID-19 alongside ground truth labels of \textit{fake claim} and \textit{factual claim} manually assigned by human coders.
We leveraged $484$ \textit{fake claim} tweets (e.g., ``only older adults and young adults are at risk’’) and $8,092$ \textit{factual claim} tweets (e.g., ``5G mobile networks do not spread COVID-19’’) tweeted by the WHO official account. 
}

\medskip \noindent
\changed{
\textbf{Dataset 4.} Paka et al.~\cite{paka2021cross} released the \underline{C}OVID-19 \underline{T}witter \underline{f}ake news (CTF) dataset\footnote{\url{https://github.com/williamscott701/Cross-SEAN}}, which consists of a mixture of labeled and unlabeled tweets pertaining to COVID-related narratives. 
As we are interested in data with ground truth labels, we disregarded the unlabeled portion of this dataset and focused only on its labeled part, which comprises $45,261$ tweets, wherein $18,555$ are labeled as \textit{genuine} and $26,706$ as \textit{fake}. 
To compose this portion of the dataset, the authors collected the tweet IDs from two other publicly available Twitter datasets of general COVID-19 narratives, and their own set of COVID-related tweets from several governmental health organizations' accounts (WHO, CDC, NIH, CPHO, PHE, and HHS) using predefined keywords such as \textit{covid19}, \textit{wuhan}, and \textit{bioweapon}. 
Next, the tweets were automatically labeled based on two main assumptions: tweets including URLs for fake news articles are fake tweets while those containing URLs for factual news articles are genuine tweets, and
tweets collected from the aforementioned health organizations are all genuine. 
This resulted in $5.3$K tweets labeled as \textit{fake} and $17.5$K as \textit{genuine}, which the authors later pre-processed with two transformer models, BERT~\cite{reimers2019sentence} and RoBERTTa~\cite{liu2019roberta} to generate new labeled tweets. 
Thus, Paka et al. artificially increased the size of their dataset to $18,555$ \textit{genuine} and $26,706$ \textit{fake} tweets. 
Lastly, they asked three annotators with expertise in fact verification to manually verify the labels of $16,000$ randomly selected tweets (split evenly into \textit{fake} and \textit{genuine}). 
With a Krippendorf's $\alpha$ of 0.82, they found that 92\% of the automatically assigned labels matched the labels given by the human annotators. In other words, on average, their approach for automatically labeling tweets as \textit{fake} or \textit{genuine} led to 8\% of mislabeled data. 
At the time of our data collection, the dataset was not entirely available; instead, the authors released a sample of $2,000$ \textit{fake} and $2,000$ \textit{genuine} tweets, which we included in our datasets of COVID-19 \texttt{misleading} and \texttt{factual} claims, respectively.}

\medskip \noindent
\changed{
\textbf{Dataset 5.} 
This dataset\footnote{\url{https://github.com/gmuric/avax-tweets-dataset}}, released by Muric et al.~\cite{Muric2021}, contains tweets pertaining to anti-vaccine narratives, including falsehoods and conspiracies related to the COVID-19 vaccine. 
The authors collected more than 1.8M tweets containing keywords that indicate opposition to the COVID-19 vaccine (e.g., \textit{antivaccine}, \textit{covidvaccineispoison}, \textit{BillGatesBioTerrorist}), which were tweeted between October 2020 and April 2021.
They additionally collected more than 135M tweets from 70K accounts engaged in the active spreading of anti-vaccine narratives---i.e., tweets from a fixed set of accounts, which may restrict the diversity of the data. 
We therefore opted to consider only the first part of their dataset in our study, which contains anti-vaccine tweets posted by a myriad of users. Such tweets were included in our COVID-19 \texttt{misleading} claims dataset.
}

\subsubsection{General Topics Tweets}
\changed{
We combined four other sources of data to compose two diverse datasets of \texttt{misleading} and \texttt{factual} claims about general topics (e.g., politics, terrorist conflicts, entertainment, etc.).
}

\medskip\noindent
\changed{
\textbf{Dataset 6.} 
Mitra and Gilbert~\cite{mitra2015credbank} released CREDBANK, a large-scale crowd-sourced dataset of approximately 60M tweets covering 96 days starting from October 2014. 
All tweets were related to $1,049$ real-world news events, each analyzed for credibility by $30$ annotators from Amazon Mechanical Turk. 
The authors continuously collected a 1\% sample of global tweets, removing spam and non-English tweets. 
For every 1M tweets, an online LDA model was used to separate a collection of tweets into sets of coherent topics, where each topic can be interpreted by the top $N$ terms. 
Each event was rated on a 5-point Likert scale of truthfulness, ranging from \textit{certainly inaccurate} to \textit{certainly accurate}, achieving a fairly high agreement score (Average Random Raters ICC = $0.77$; 95\% C.I. = $[0.77, 0.81]$).
We opted to disregard tweets labeled as \textit{probably inaccurate}, \textit{uncertain (doubtful)}, and \textit{probably accurate} as they likely mixed ambiguous claims. 
However, we were unable to identify events rated as \textit{certainly inaccurate} by all annotators, and hence decided to not fetch tweets belonging to this category. 
Therefore, we selected 18 events which were rated \textit{certainly accurate} by all 30 annotators, for a total of $1,943,827$ tweets.
}

\medskip\noindent
\changed{
\textbf{Dataset 7.} 
The Russian Troll Tweets Kaggle dataset\footnote{\url{https://www.kaggle.com/vikasg/russian-troll-tweets?select=tweets.csv}} contains $200$K tweets from malicious accounts connected to Russia’s Internet Research Agency (IRA) posted between July 2014 and September 2017. This dataset was reconstructed by a team at NBC News\footnote{\url{https://www.nbcnews.com/tech/social-media/now-available-more-200-000-deleted-russian-troll-tweets-n844731}} after Twitter deleted data from almost $3$K accounts believed to be connected with the IRA in response to an investigation of the House Intelligence Committee into how Russia may have influenced the 2016 U.S. election.
}

\medskip\noindent
\changed{
\textbf{Dataset 8.} 
Vo and Lee~\cite{vo2019learning} released a dataset\footnote{\url{https://github.com/nguyenvo09/LearningFromFactCheckers}} of tweets that were fact-checked based on news articles from two popular fact-checking websites (Snopes and Politifact). 
The authors originally collected $247,436$ fact-checked tweets posted between May 2016 through 2018. 
Because the data collection took place during the 2016 U.S. presidential election, many tweets contain misinformation related to then-candidates Hillary Clinton and Donald Trump. 
The authors discarded tweets that had been removed by Twitter, as well as non-English tweets and tweets linking to news articles that could not be classified as entirely true or false.
After that, their final dataset consisted of $73,203$ fact-checked tweets, where $59,208$ were labeled as \textit{fake} and $13,995$ as \textit{true}, which we included in our datasets of \texttt{misleading} and \texttt{factual} claims, respectively.
}

\medskip\noindent
\changed{
\textbf{Dataset 9.} 
This dataset\footnote{\url{https://shanjiang.me/resources/\#misinformation}} was released by Jiang et al~\cite{jiang2018linguistic} who collected a variety of user comments to analyze whether linguistic signals could be used to detect misinformation in posts or comments. 
For their data collection, the authors crawled through all officially fact-checked articles from the Politifact and Snopes archives up until January 2018. 
The portion of their dataset exclusively from Twitter originally contained a total of $2,327$ tweets along with ground truth labels across a spectrum of fact-check ratings. 
Tweets were labeled as either \textit{true}, \textit{mostly true}, \textit{half true}, \textit{mostly false}, \textit{false}, and \textit{pants on fire} (i.e., when the statement was not accurate and made a ridiculous claim). 
As we are specifically interested in purely misleading and factual claims, we included only \textit{true} ($N=231$) in our dataset of \texttt{factual} claims, and both \textit{false} ($N=1130$) and \textit{pants on fire} ($N=134$) tweets in our dataset of \texttt{misleading} claims. 
}

\subsection{Data Collection \& Stratified Random Sampling}
\changed{
First, we discarded repeated tweet IDs from the four composed datasets. 
We then used the Twitter API to collect these tweets along with metadata related to the tweets themselves (e.g., language, lists of hashtags, symbols, user mentions, and URLs included), the users/authors of the tweets (e.g., name, profile description, account date of creation, number of followers, number of friends), and the tweet engagement (e.g., number of retweets and number of likes). 
However, we were able to retrieve only a portion of tweets per each dataset. 
Many tweets were no longer available/accessible by the time of the data collection (especially those containing misleading claims), most likely because they had been deleted by either Twitter or the user. 
Moreover, we discarded non-English language tweets and tweets containing no text or very short texts. 
Upon collecting the entire dataset, we dropped $416,283$ entries with null values for the combined engagement metric---this likely was due to errors during poor parsing of the json strings after collecting the entire datasets; nonetheless, this step left us with $2,116,397$ total tweets (summarized in  Table~\ref{tab:dataset_summary}).
}

\begin{table}[]
\centering
\caption{\changed{Descriptive statistics of our final four datasets based on the combined engagement metric.}}
\label{tab:dataset_summary}
\resizebox{0.75\textwidth}{!}{%
\begin{tabular}{@{}crrrrr@{}}
\toprule
         & \multicolumn{2}{c}{\textbf{Factual}}    &  & \multicolumn{2}{c}{\textbf{Misinformation}} \\ \cmidrule(lr){2-3} \cmidrule(l){5-6} 
         & \textit{COVID-Related} & \textit{General Topics} &  & \textit{COVID-Related} & \textit{General Topics} \\ \midrule
$N$         & 9,111         & 1,243,913      &  & 828,501       & 32,243         \\
$n_{strata}$ & 4,814        & 4,448          &  & 4,533         & 4,147          \\   
$\mu$      & 368.5         & 9,791.6        &  & 2,214.3       & 3,014.7        \\
$\sigma$   & 7,157.9       & 73,305.6       &  & 10,051.9      & 28,727.4       \\ 
\textit{Mean Rank}  & 2407.5        & 2244.5         &  & 2267.0        & 2074.0          \\ \bottomrule
\end{tabular}%
}
\end{table}

\begin{figure}
    \centering
    \includegraphics[width=\textwidth,trim={1.5cm 1.5cm 2cm 1.4cm},clip]{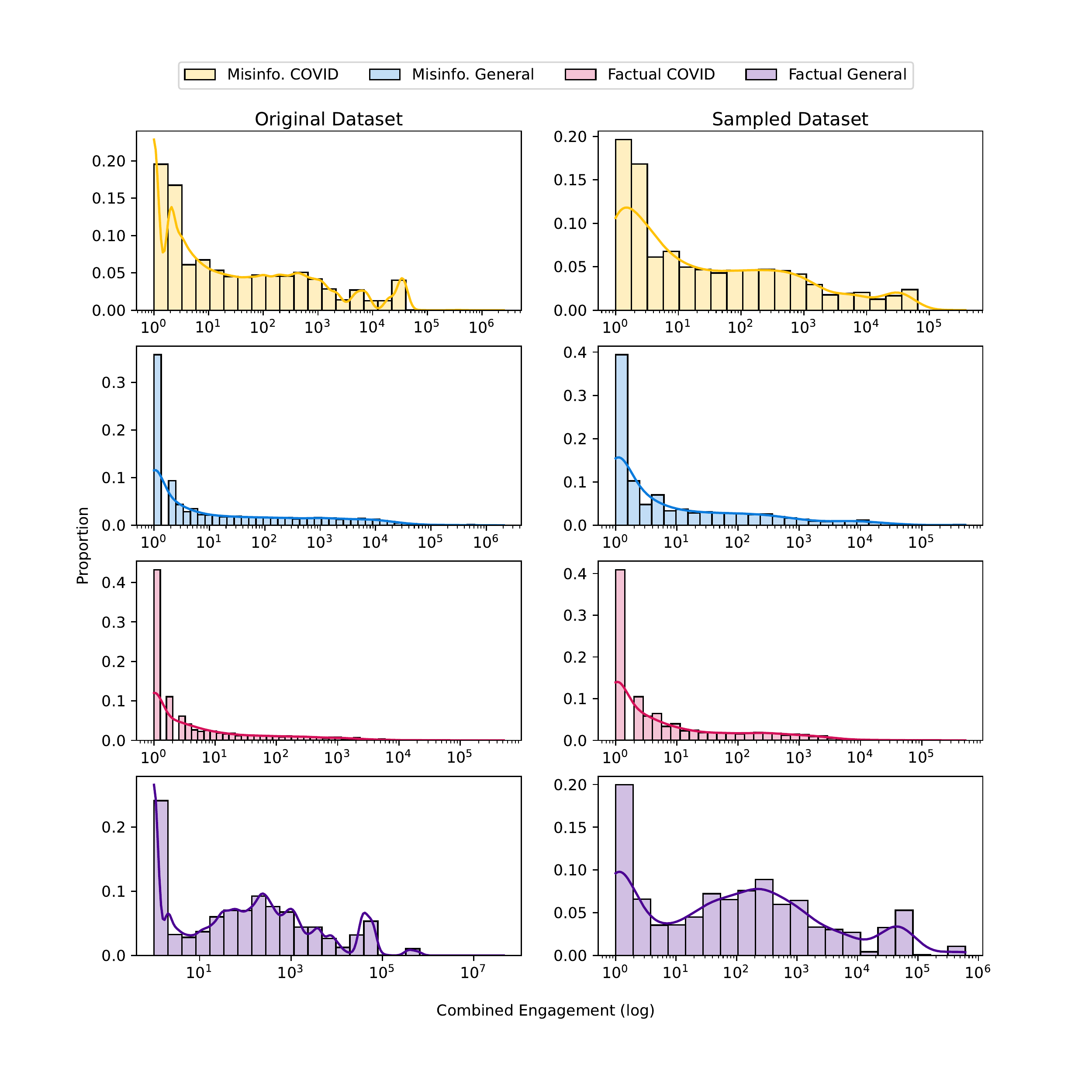}
    \caption{\changed{Comparison of the distribution of combined engagement (log-normalized) versus the proportion of its occurrence for each dataset.}}
    \vspace{-1.5em}
    \label{fig:datasets_distribution}
\end{figure}

\changed{
Although this data imbalance is not ideal for statistical analyses as it can introduce biases into our results, we argue that this imbalance is unfortunately part of the phenomenon of misinformation itself; COVID-related tweets, for example, are laden with misinformation given that the scientific research on this topic is constantly evolving. 
Such an environment of uncertainty facilitates rumors and misinformation~\cite{Allport1947-on}.
In order to appropriately decrease the imbalance between the datasets, we opted to perform stratified random sampling to obtain sample populations that adequately represented the distribution of the combined engagement metric within each dataset. In other words, instead of randomly selecting data from each of the four datasets, we sampled  subgroups, i.e., \textit{strata}, of $n \approx 4,556$ from each dataset according to the distribution of combined engagement. This $n$ was chosen as it is 50\% of the smallest population size across our datasets (i.e., $N = 9,111$ for COVID-related factual tweets), and allowed us to maintain variability across all class sizes. Inspired by $k$-folds commonly employed in machine learning, we repeated this process $10$ times, obtaining a total of 10 stratified random samples of $17,982$ total tweets each. Figure~\ref{fig:datasets_distribution} compares the distribution of the original datasets with one of the stratified random samples, showing that we did indeed stay true to the original distribution of combined engagement from our original datasets. 
}

\subsection{Data Preprocessing}\label{sec:data_preprocessing}
Before feature extraction, the \changed{full text of the collected tweets were preprocessed by removing numbers (e.g., ``1 million’’ or ``12,345’’ become `` million’’ and ``,’’), emojis, hashtags (e.g., ``\#COVID’’), mentions (e.g., ``@WHO’’), and URLs.} 
Other typical NLP preprocessing steps, such as tokenization, removal of stop words, and lemmatization were not performed, as both LIWC and sentiment analysis packages can work with raw text. 

\subsection{Feature Extraction}
From the cleaned dataset, we extracted a total of \nfeatures features per tweet, including features derived from the metadata (i.e., tweet- and user-related descriptors), addressing sociolinguistic (e.g., cognitive and structural components, such as formal and logical language) \changed{and moral frames (e.g., fairness or reciprocity)}, as well as sentiment characteristics of the tweet texts.

\subsubsection{Tweet Metadata, User Metadata, and Engagement} \label{sec:features_metadata_eng}
We extracted the following features from the collected Twitter metadata:

\begin{itemize}[leftmargin=*]
\item \changed{Six} {\bf tweet-related} features: \# of likes, \# of retweets, \# of hashtags, \# links/URLs, \changed{\# of combined engagement (i.e., \# retweets + \# likes)}, and \# of emojis in the tweet.
\item \changed{Twelve} {\bf user-related} features: \# of followers, \# of friends, \# of lists,  \# of favorited tweets, \changed{verified (binary), presence of profile image (binary), use of default profile image (binary) or default profile (binary), whether geolocation is enabled (binary), whether the user has an extended profile (binary) or background tile (binary), and \# of tweets made by the user}.
\end{itemize}

We included the combined engagement metric in our analysis: \# of likes plus \# of retweets. Note that this engagement metric was log-normalized following \changed{Aldous and Jansen} \cite{aldous2019view} because it was highly skewed, with more weight in the left tail of the distribution. 
According to the authors, engagement can be captured in a 4-level scale where lower levels indicate fewer public expressions of engagement. 
Likes and retweets encompass different levels of user engagement, with the former being classified as level-2 and the latter as level-4 (the highest level). 
The reasoning behind this is intuitive: while liking a tweet does involve some degree of engagement because the user publicly exposes their preference about a certain topic, retweeting a given content displays a higher level of engagement because the user deliberately seeks to amplify the reach of that content by disseminating it through different networks using their own feed---in other words, a retweet is the most public level of engagement that can be observed for Twitter. The most private level of engagement (level-1) includes viewing content posted by other users, while commenting on content is considered level-3 as it involves more public exposure than simply liking it but less exposure than sharing (retweeting) it.
As the collected dataset does not have level-1 and level-3 engagement metrics, we decided to analyze likes (level-2) and retweets (level-4) \changed{as a single combined metric}.

Additionally, because emojis (pictures) and emoticons (icons created with marks, letters, and numbers, e.g., ``:-)'') are commonly used in tweets to display sentiment or emotions, we opted to implement an emoji and emoticon counter. However, we later decided to disregard emoticons after observing a high occurrence of false positives; many of the combinations of regular punctuations were incorrectly identified as emoticons (e.g., in tweets containing the text ``d:'' and ``\%)'' were misclassified as emoticons). In other words, we only counted for emojis.

\subsubsection{Sociolinguistic Analysis}
We performed a sociolinguist analysis on the collected tweets using the Linguistic Inquiry and Word Count (LIWC) software (version 2015)~\cite{pennebaker2015development}. This tool estimates the rate at which certain emotions, moods, and cognition (e.g., analytical thinking) are present in a piece of text based on word counts (e.g., the words ``nervous,'' ``afraid,'' and ``tense'' are counted as expressing anxiety). More specifically, we extracted 93 features related to emotional, cognitive, and structural components from the collected tweets, including:

\begin{itemize}[leftmargin=*]
\item Four \textbf{language metrics}: total number of words, average number of words per sentence, number of words containing more than six letters, and number of words found in the LIWC dictionary.
\item Eighty-five \textbf{dimensions}, including function words (e.g., pronouns, articles, prepositions), grammar characteristics (e.g., adjectives, comparatives, numbers), affect words (e.g., positive and negative emotions), social words (e.g., family, friends, male/female referents), cognitive process (e.g., insight, certainty), core needs (e.g., power, risk/prevention focus), time orientation (e.g., past/present/future focus), personal concerns (e.g., home, money, death), informal speech (e.g., swear words, netspeak), and punctuation (e.g., periods, commas, question marks). These features reflect the percentage of total words per dimension (e.g., ``positive emotions'' equal to 7.5 means that 7.5\% of all words in the tweet were positive emotion words).
\item Four \textbf{summary variables} expressed in a scale ranging from 0 (very low) to 100 (very high): (i) analytical thinking, which is related to formal, logical, and hierarchical thinking patterns based on eight different word dimensions; (ii) clout, which refers to relative social status, confidence, or leadership expressed through writing; (iii) authenticity, which refers to writing that is personal and honest, and; (iv) emotional tone, where the higher the number, the more positive and upbeat the tone (scores lower than 50 usually suggest a more negative tone).
\end{itemize}

\subsubsection{Moral Frames Analysis}
\changed{
We measured moral frames using the moral foundations dictionary~\cite{Graham2009-ns} dictionary in LIWC. Based on moral foundations theory~\cite{Haidt2007-hm}, the authors aggregated 295 words for each five of moral intuitions encompassing 11 total features, which encompass psychological preparations for reacting to issues pertaining to \textbf{harm/care}, \textbf{fairness/reciprocity}, \textbf{ingroup/loyalty}, \textbf{authority/respect}, and \textbf{purity/sanctity}.
}

\subsubsection{Sentiment Analysis}
\changed{For comprehensiveness}, we considered three out-of-the-box packages for the sentiment analysis of our dataset: 

\begin{enumerate}[leftmargin=*]
\item\textbf{VADER}~\cite{gilbert2014vader} is a rule-based NLP library built specifically for sentiment analysis and is available with NLTK, a popular NLP suite of libraries. Among the outputs provided by VADER is the \textit{compound score}, a uni-dimensional normalized, weighted composite score. A compound score $\geq 0.05$ denotes a positive sentiment, between $-$0.05 and 0.05 denotes a neutral sentiment, and $\leq -0.05$ denotes a negative sentiment.
\item\textbf{TextBlob}~\cite{loria2014textblob} is another rule-based NLP library and offers the \textit{polarity} score, which ranges from $-$1.0 (most negative sentiment) to 1.0 (most positive sentiment). Inspired by VADER’s compound score system, we opted to label scores between $-$0.05 and 0.05 as neutral sentiment.
\item\textbf{Flair}~\cite{akbik-etal-2018-contextual} is an embedding-based framework built on PyTorch.  Its pre-trained sentiment models output value as \texttt{positive} or \texttt{negative} along with the confidence score in the range of 0.5 (low confidence) to 1.0 (high confidence). To add a new sentiment label \texttt{neutral}, we normalized the confidence scores in the range of 0.0 to 1.0, and negated the confidence score for negative sentiment. Following VADER’s compound score system, we then labeled scores between $-$0.05 and 0.05 as neutral sentiment.
\end{enumerate}

\vspace{-0.2em}

We tested the above three sentiment analysis packages with a dataset of 44,957 COVID-relevant tweets, manually labeled for sentiment ($19,592$ positive, $17,031$ negative, and $8,334$ neutral)~\cite{covid_labelled_tweet_dataset} using the F-score as the performance metric. We chose VADER for our sentiment analysis because it outperformed not only the two other sentiment analysis packages, but also all three packages combined in a voting scheme (average F-score of 82.0\% with VADER versus 48.0\% with Flair, 50.0\% with TextBlob, and 77.0\% with the voting scheme). Then, for each collected tweet, we extracted three binary sentiment features: positive, negative, and neutral.

\subsection{Correlation Analysis}

\changed{
To address RQs 3 and 4 pertaining to the correlation of engagement with COVID- and non-COVID-related misinformation and factual tweets, we 
measured feature importance based on Pearson’s correlation coefficient, $r$. However, aware that $r$ only adequately describes linear relationships, we have opted to expand our analyses by additionally finding the fixed point of Maximal Correlation (MC) using the Alternating Conditional Expectations (ACE) algorithm on each feature. 
This process involves transforming all variables in such a way as to maximize $r$ for the dependent and independent variables. 
This process has been shown to be robust against noisy data and several different relationship types, allowing for non-linear correlations to be more accurately detected compared to Pearson’s correlation coefficient~\cite{Deebani2018-rt}. 
Note that Maximal Correlation ranges from $[0,1]$ and therefore does provide information regarding the polarity of the correlation. 
}

\changed{
Additionally, the Pearson correlation coefficient is biased such that the simple mean of $r$ of all 10 samples would underestimate the true $r$. 
Therefore, performing a Fisher $z$-transformation correction of the $r$s allows us to reduce bias and more accurately estimate the population correlation~\cite{Corey1998-ce}. 
In other words, we report the average Pearson’s correlation coefficient, $r_{z}$, i.e., the inverse $z$-transform of the averaged $z$-values over all 10 samples. Additionally, to combine the $p$-values obtained for each sample into a single metric, we rely on the Fisher method by sum of logs.
}

\section{Results}\label{sec:analysis_results}
This section details the statistical and \changed{correlation} analyses performed on \changed{our curated} dataset to answer each of our \changed{four} research questions, as well as their results. All statistical tests were performed based on a 1\% significance level ($\alpha = .01$).
\changed{
Tables~\ref{tab:stats_results_engagement} and \ref{tab:correlation_analysis} summarize our results.
}

\subsection{RQ1: \changed{Are COVID-19 misinformation tweets more engaging than COVID-19 factual tweets?}}
To address this research question, we compared the log-normalized combined engagement metric between the factual and misinformation tweets. More specifically, we first confirmed that \changed{the combined engagement metric across the stratified random samples} does not follow a normal distribution for the factual and misinformation groups of tweets using the Shapiro-Wilk ($p<.001$) and D’Agostino’s K-squared ($p<.001$) tests. 
Note that \changed{we observed that most tweets were heavily skewed at near-zero, likely a naturally occurring phenomenon of tweets}. 
We also confirmed that the distribution of the data was not homogeneous by comparing the log-normalized combined engagement metric for the factual tweets group against the misinformation tweets group \changed{($W = 378.89$, $p<.001$)}. Therefore, we opted to analyze the log-normalized combined engagement metric for the two groups of tweets using non-parametric tests. 
\changed{Note, too, that these results were nearly exact for all stratified random samples, again indicating that the strata adequately reflected the distribution of combined engagement.}
From the Two-Sample Kolmogorov Smirnov test, we determined that the distribution of combined engagement for COVID-related factual tweets is significantly different from that of the COVID-related misinformation tweets \changed{($KS = 0.21, p<.001$)}. 

\begin{table}[]
\centering
\caption{Summary results for statistical tests conducted on engagement metrics and bot/user account labels.}
\label{tab:stats_results_engagement}
\resizebox{\textwidth}{!}{%
\begin{tabular}{@{}rcrl@{}}
\toprule
\multicolumn{1}{c}{\textbf{Data}} &
  \textbf{Measure} &
  \multicolumn{2}{c}{\textbf{Measurement Statistics}} \\ \midrule
\begin{tabular}[c]{@{}r@{}}Combined Engagement \\ (raw)\end{tabular} &
  Shapiro-Wilk &
  \multicolumn{1}{c}{\begin{tabular}[c]{@{}c@{}}Factual COVID-Related\\ Misinformation General Topics\\ Factual General Topics\\ Misinformation General Topics\end{tabular}} &
  \begin{tabular}[c]{@{}l@{}}$W=0.7875$***\\ $W=0.8946$***\\ $W=0.9374$***\\ $W=0.7969$***\end{tabular} \\ \midrule
\multirow{3}{*}{\begin{tabular}[c]{@{}r@{}}\\ Combined Engagement\\ (log-norm)\end{tabular}} &
  Levene &
  \begin{tabular}[c]{@{}r@{}}Factual vs. Misinformation COVID-Related\\ Factual vs. Misinformation General Topics\end{tabular} &
  \begin{tabular}[c]{@{}l@{}}$W=378.89$***\\ $W=359.59$***\end{tabular} \\ \cdashline{2-4}
 &
  Two-Sample Kolmogorov-Smirnov &
  \begin{tabular}[c]{@{}r@{}}Factual vs. Misinformation COVID-Related\\ Factual vs. Misinformation General Topics\end{tabular} &
  \begin{tabular}[c]{@{}l@{}}$K_2 = 0.2133$***\\ $K_2 = 0.3459$***\end{tabular} \\  \cdashline{2-4}
 &
  Mann-Whitney U &
  \begin{tabular}[c]{@{}r@{}}Factual vs. Misinformation COVID-Related\\ Factual vs. Misinformation General Topics\end{tabular} &
  \begin{tabular}[c]{@{}l@{}}$U=7,662,279$***, $r = 0.35$\\ $U=5,725,193$***, $r= 0.31$\end{tabular} \\ \bottomrule
\multicolumn{3}{l}{\begin{tabular}[c]{@{}l@{}}*** Significant at  $p < .001$\\\end{tabular}}
\end{tabular}%
}
\end{table}

\changed{
A comparison of the mean distribution of factual and misinformation tweets was desirable, given the notable differences in the overall populations ($\mu_{COVID, factual} = 368.5$ and $\mu_{COVID, misinfo} = 2,214.3$. 
However, due to the non-normality and skew of these variables, we opted to conduct the Mann-Whitney U-test and compare the mean ranks of the two samples.
For each strata, factual COVID-19 tweets ($n = 4,814$) had a larger average mean rank ($2,407.5$) than misinformation tweets ($n = 4,533, \mu_{rank}=2,267.0$). 
Therefore, we find that 
}
the combined engagement of the factual tweets was statistically and significantly higher compared to the misinformation tweets \changed{$U = 7,662,279$, $p<.001$), indicating that factual COVID-19 tweets tend to be more engaging than COVID-19 misinformation tweets.} 
Given that $U_{max} = n_{strata, 1} \times n_{strata, 2} = 21,821,862$, we can convert the U-statistic to an effect size, $r = U/U_{max} = 0.35$. Put into words, there is a \changed{medium} probability that a combined engagement value from the factual tweets will be greater than misinformation tweets. 

\smallskip
\noindent
\begin{tcolorbox}
Misinformation tweets about COVID-19 were statistically and significantly less engaging than \changed{COVID-19} factual tweets.
\end{tcolorbox}

\subsection{RQ2: \changed{Are general topic misinformation tweets more engaging than general topic factual tweets?}}
\changed{
We repeated the analyses conducted for \textbf{RQ1}, finding that the combined engagement metrics also do not follow normal distribution based on the Shapiro-Wilk ($p<.001$) and D’Agostino’s K-squared ($p<.001$) tests, and that the distribution of the data was not homogeneous for the two groups ($W = 359.59$, $p<.001$).
The Two-Sample Kolmogorov Smirnov test also showed that the distribution between factual and misinformation general topic tweets were significantly different ($KS = 0.35, p <.001$). 
}

\changed{
The Mann-Whitney U-test similarly showed that factual general topic tweets ($n= 4,448$) had a larger average mean rank ($2244.5$) than misinformation tweets ($n = 4,147, \mu_{rank} = 2,074$).
Therefore, the combined engagement of the general topic factual tweets was also significantly higher compared to the misinformation tweets ($U = 5,745,193.0, p<.001$), showing that factual general topic tweets tend to me more engaging than misinformation tweets.
We then convert this U-statistic to an effect of $r=0.31$, indicating that there is also a medium probability that a combined engagement value from the factual general topic tweets will be greater than the misinformation tweets. 
}

\smallskip
\noindent
\begin{tcolorbox}
\changed{Misinformation tweets about general topics were statistically and significantly less engaging than factual tweets.}
\end{tcolorbox}

\begin{table}[]
\centering
\caption{\changed{Summary of correlation analysis between the log normalized combined engagement metric and all features. Only $r_{z}$ values indicating a moderate correlation ($>0.5$) and with a combined MC $p$-value $<.01$ are shown.}}
\label{tab:correlation_analysis}
\resizebox{0.7\textwidth}{!}{%
\begin{tabular}{@{}clcc@{}}
\toprule
\multicolumn{1}{l}{\textbf{Feature Type}} & \textbf{Feature} & \textbf{$r_z$} & \textbf{(MC) $r_z$} \\ \midrule
\multicolumn{4}{l}{\textbf{Factual: COVID-Related}}                                              \\ \hdashline
\multirow{20}{*}{\textit{LIWC}}         & Affective Processes                     & 0.53  & 0.71 \\
                                        & All Punctuation                         & -0.05 & 0.58 \\
                                        & Assent (Informal Language)              & 0.65  & 0.74 \\
                                        & Clout                                   & 0.36  & 0.56 \\
                                        & Colon (Punctuation)                     & 0.34  & 0.54 \\
                                        & Dictionary Words                        & 0.13  & 0.56 \\
                                        & Past Focus                              & 0.49  & 0.66 \\
                                        & Informal Speech                         & 0.62  & 0.72 \\
                                        & Insight (Cognitive Processes)           & 0.32  & 0.68 \\
                                        & Male Referents (Social Words)           & 0.77  & 0.88 \\
                                        & Netspeak (Informal Language)            & 0.66  & 0.77 \\
                                        & Positive Emotion (Affect Words)         & 0.52  & 0.78 \\
                                        & Person Pronouns (Linguistic Dimensions) & 0.31  & 0.56 \\
                                        & Question Marks (All Punctuation)        & -0.31 & 0.53 \\
                                        & Reward (Drives)                         & 0.33  & 0.67 \\
                                        & Sad (Affect Words)                      & 0.48  & 0.65 \\
                                        & 3rd Person Singular (Function Words)    & 0.81  & 0.91 \\
                                        & Words > 6 Letters                       & -0.26 & 0.59 \\
                                        & Social Words                            & 0.41  & 0.63 \\
                                        & Time (Relativity)                       & 0.21  & 0.51 \\
\textit{Sentiment}                      & VADER Compound                          & 0.19  & 0.66 \\ \midrule
\multicolumn{4}{l}{\textbf{Factual: General Topics}}                                             \\ \hdashline
\multirow{5}{*}{\textit{LIWC}}          & Assent (Informal Speech)                & 0.36  & 0.68 \\
                                        & Colons (All Punctuation)                & 0.20  & 0.52 \\
                                        & Informal Speech                         & 0.29  & 0.62 \\
                                        & Netspeak (Informal Speech)              & 0.32  & 0.63 \\
                                        & Prepositions (Function Words)           & 0.02  & 0.54 \\ \midrule
\multicolumn{4}{l}{\textbf{Misinformation: COVID-Related}}                                       \\ \hdashline
\multirow{7}{*}{\textit{LIWC}}          & Assent (Informal Speech)                & 0.26  & 0.75 \\
                                        & Colons (All Punctuation)                & 0.34  & 0.75 \\
                                        & Informal Speech                         & 0.19  & 0.69 \\
                                        & Impersonal Pronouns                     & 0.06  & 0.64 \\
                                        & Netspeak (Informal Speech)              & 0.26  & 0.73 \\
                                        & Quotation Marks (All Punctuation)       & 0.10  & 0.50 \\
                                        & Word Count                              & -0.10 & 0.51 \\ \midrule
\multicolumn{4}{l}{\textbf{Misinformation: General Topics}}                                      \\ \hdashline
\multirow{3}{*}{\textit{User Metadata}} & Followers Count                         & 0.28  & 0.73 \\
                                        & Listed Count                            & 0.30  & 0.66 \\
                                        & User Verified                           & 0.53  & 0.53 \\ \bottomrule
\end{tabular}%
}
\end{table}

\subsection{RQ3: \changed{Which features are most correlated with engagement in COVID-19 vs. general topics misinformation tweets?}}

\changed{
Our correlation analysis revealed that few of the extracted features were strongly correlated ($r_{MC,z} \geq 0.5$) with the log-normalized combined engagement metric. 
COVID-related misinformation combined engagement was notably strongly correlated with LIWC-based grammar features (i.e., use of informal speech, punctuation, impersonal pronouns) and word count (ranging from $[0.50, 0.75]$).
General topics misinformation stood in contrast to all other groups in that only three features showed strong correlation, all related to user metadata: number of followers ($r_{MC, z} = 0.73$), the number of public lists of which that a user is a member  ($r_{MC, z} = 0.66$), and whether the user is verified ($r_{MC, z} = 0.53$).
}

\smallskip
\noindent
\begin{tcolorbox}
\changed{
The top features related to engagement for COVID-19 and general topics misinformation were, respectively, the tweet's grammar (e.g., use of informal speech) and user metadata (e.g., verified user).
}
\end{tcolorbox}

\subsection{RQ4: \changed{Which features are most correlated with engagement in COVID-19 vs. general topics factual tweets?}}

\changed{
Factual COVID-related tweets contained several strong correlations relative to the remaining groups. 
Among the many grammar-related features (e.g., punctuation, informal language, etc.) present, the highest correlation ($r_{MC,z} = 0.91$) occurred with the use of third-person singular words (e.g., \textit{she, her, him}), a feature not strongly correlated with any other group; the second-highest correlation ($r_{MC,z} = 0.88$) was similarly related to male referents (e.g., \textit{boy, his, dad}). 
Also in contrast with other groups, factual COVID tweets were strongly correlated with effective processes ($r_{MC,z} = 0.71$) and emotion, as measured both by LIWC ($r_{MC,z, positive} = 0.78$ and $r_{MC,z, sad} = 0.65$) and VADER ($r_{MC,z} = 0.66$). 
Clout ($r_{MC,z} = 0.71$) was the only LIWC summary variable to appear among any of the groups and indicates confidence and leadership in writing.
Factual general topics tweets, however, only strongly correlated with LIWC's grammar features (i.e., informal speech, punctuation, and prepositions), similarly to the strongly correlated features for COVID-related misinformation.
}

\smallskip
\noindent
\begin{tcolorbox}
\changed{
The top features related to engagement for COVID-19 factual tweets pertained to grammar (e.g., use of netspeak), emotion (both positive and negative), and the writer's confidence, whereas general topic tweets pertained solely to grammar (e.g., use of colons or prepositions).}
\end{tcolorbox}

\section{Discussion}\label{sec:discussion}
In this paper, we set out to answer \changed{four} research questions relating COVID-19 \changed{and general topics} tweets as a function of \changed{combined engagement metric}. 
This section summarizes the take-aways and limitations of our work, and suggests possible future directions of research. 

First, it is important to note that distinguishing between factual and misinformation tweets is challenging as research has shown that automatic detection of misinformation is a nuanced and open research problem in the machine learning field~\cite{Zhou2020-fo} and social media platforms are inherently rooted in big data that is unstructured and noisy~\cite{Shu2017-pw}. Such problems exacerbate the difficulty of detecting misinformation.
The digital revolution and the integration of social media into our daily lives have been leveraged as tools for the faster propagation of disinformation campaigns. Research has shown that humans are poor at detecting deception~\cite{Granhag2004-qj} and our ability to detect digital fake news is ``bleak''~\cite{Wineburg2016-sd}. Understanding how machines can detect highly engaging dis/misinformation will provide a first line of defense against deception in the online sphere. This knowledge can be used by government agencies and organizations to convey critical public health information to the general populace. For example, with respect to the Italian Ministry of Health, Lovari~\cite{lovari2020spreading} found that keeping the public constantly informed via dissemination of information in understandable forms (e.g., data and visuals), helps reduce the spread of misinformation.  

\changed{
Therefore, a primary purpose of this work was to point researchers towards potentially impactful metadata that could give inklings towards purposeful or unintentional false information.
Importantly, we found that misinformation tweets pertaining to general topics strongly correlated with the users' metadata; these features all contained a positive polarity in terms of $r_z$, potentially indicating that influential users were responsible for generating engagement with general topics misinformation.
}
Operating under the assumption that a real Twitter account is more likely to be verified and \changed{have several followers}, we can infer that \changed{misinformation} tweets by seemingly real \changed{and influential} users offer \changed{a perceived sense of credibility towards the user's claims}. 
This corroborates prior research providing evidence that the credibility of the source of the information greatly helps in the detection of fake news~\cite{Zhou2020-fo}\changed{---however, when applied to the context of misinformation, may prove the opposite and be even more deceiving to the average user.}

As such, the \changed{semantic} content of the tweet itself (based on LIWC analysis) appears not to be relevant for engagement \changed{
(except for factual COVID tweets). Instead, the \textit{syntax} of the tweet was highly correlated with engaging tweets for factual COVID tweets, and both factual and misinformation general topics.
} 
Interestingly, we found that tweet sentiment was not relevant to predict engagement, \changed{except in the context of truthful COVID-related tweets.}

\changed{
In stark contrast to this, we found that COVID-related factual tweets were the only group that showed inklings towards the importance of semantic content.
More specifically, engagement was strongly correlated with sentiment and cognitive processing-related keywords, potentially indicating that engaging factual tweets in the context of COVID-19 were those that appealed to one's pathos. 
Tweets with fewer complex words (i.e., $> 6$ letters) and question marks also showed a negative Pearson's correlation, $r_z$, and strong MC correlation ($r_{MC,z} = 0.59$ and $0.53$, respectively), likely pointing towards the presence of understandable and declarative statements. 
Our results therefore indicate that misinformation is not engaged with equally.
Understanding and mitigating different types of misinformation (e.g., political, health, science) might be better done in a targeted, per-issue fashion, rather than as a lumped and ill-fitting one-size-fits-all manner. 
}

We also found that factual tweets were statistically more engaging than misinformation tweets, \changed{regardless of the context of the tweet (general topics or COVID-19)}. 
To the best of our knowledge, our study is the first to analyze engagement in COVID tweets \changed{relative to veracity and other topics}. 
\changed{Surprisingly, we did not find that the \# of ULRs in the tweet was a strongly correlated feature}. We suspected that URLs could potentially increase the veracity of the information presented in the tweet, thus helping distinguish factual information from misinformation and reinforce false claims in misinformation tweets, therefore increasing their engagement.

\subsection{Limitations \& Future Works} \label{sec:dicussion_limitations_future}

\subsubsection{\changed{Dataset imbalance} and representativeness} 
Our \changed{original} dataset was highly imbalanced, wherein \changed{factual general topics and COVID-related misinformation accounted for a much larger portion than factual COVID-related and general misinformation tweets.}
Though we see this limitation as part of the phenomenon, \changed{
we took steps to ameliorate this by generating $10$ unique stratified random samples of near-equal sizes.
It is notable, however, that factual COVID tweets exhibited many more strong correlations and variety compared to the three remaining groups. Though we sample 50\% of this group to introduce variability, it is undoubtedly inevitable that there were repeated tweets across the strata, thus strengthening correlations.
}
This lack of variety of \changed{factual COVID tweets} data limits the generalizability of our findings \changed{pertaining to factual COVID-related tweets}. 
In addition, this can potentially affect the performance of \changed{future} machine learning models attempting to distinguish factual from misinformation tweets, likely causing the models to overfit the factual tweets (the group with most \changed{repeated samples across the $k$-folds}). 
A possible way to overcome this limitation of imbalanced data---potential advice for future works---would be to increase the number of \changed{factual COVID} tweets artificially, by generating larger synthetic datasets applying GAN~\cite{shamsolmoali2020imbalanced}, \changed{adopting a downsampling strategy, or decreasing the size of the strata and increasing the number of $k$-folds.}

\changed{
Additionally, our study was a meta-analysis covering nine datasets. 
Although our final sample size of 2.1M is relatively large, we are cognizant that we were unable to collect several tweets from across these nine datasets---these tweets could quite possibly be misinformation tweets, especially those with high engagement, removed by Twitter itself.
Though this could skew our results to favor high-engagement tweets that have not been removed by Twitter (i.e., likely factual tweets), this nonetheless might also reflect Twitter's increased ability or effort to reduce the amount of misinformation. 
As we do not have access to the removed tweets to confirm this hypothesis, we urge future work to consider a time-series analysis of tweets, following tweets from their inception up until their removal and analyzing how, if at all, engagement and veracity are contributing factors to this.
}

\changed{
And while we have taken the first steps towards demonstrating the interplay between veracity and tweet context, showing that engagement is impacted by different meta features, future work can nonetheless take a more nuanced look. 
For example, via comparison of COVID and measles vaccine hesitancy, or specific events and controversies marred in misinformation, such as the 2020 U.S. General Election. 
}

\subsubsection{Feature engineering, feature selection, and classification models}
\changed{We envision our work as a precursor to future work, especially those leveraging machine learning. While} we leveraged \nfeatures features for our data analysis, there is still a plethora of additional features that future works can investigate (see \cite{Zhou2020-fo}). 
For example, we considered only the count of emojis in a tweet, but we did not analyze emojis as a proxy of sentiment.
Works such as Novak et al.~\cite{Kralj_Novak2015-dg} and Fersini et al.~\cite{Fersini2016-bv} have found that emojis provide useful sentiment information, which may be significant in the informal communication environment of Twitter. 
Automatic feature extractors, such as Word Embedding, \changed{TF-IDF,} Word2Vec, BERT, and GloVe, may also provide relevant features to predict misinformation and engagement in tweets. 
By combining some of these feature extractors with a set of popular multi-class classifiers, Elhadad et al.~\cite{elhadad2020detecting} obtained F-scores greater than 99\% when distinguishing factual from misleading pieces of information (e.g., news) related to COVID-19. It should be noted, however, that textual feature extractors are usually sensitive to the text's size, which may be a limitation when handling tweets given their limit of characters. 
Lastly, given the size and imbalance of our dataset, \changed{future works should find ways of leveraging} state-of-the-art deep learning models for natural language processing (e.g., Recurrent Neural Networks, LSTM).

\changed{
Additionally, factual general topics tweets and misinformation COVID-related tweets contained few strongly correlated features, all encompassing grammar-related LIWC dimensions.
This indicates that future work should investigate \textit{how} tweets are written, which may be a simpler approach than attempting to fact-check each and every claim. 
}

Furthermore, works such as DiResta et al.~\cite{DiResta_IRA_report} posit that, for the 2020 US General Election cycle---which greatly overlapped and intertwined with the spread of COVID---there would be a decrease in reliance on bot accounts for the spreading of misinformation, as automation techniques have become better policed~\cite{Craig_Timberg2018-tj}. 
In fact, \changed{as mentioned above, during our collection of the tweets across the nine datasets,} we were unable to retrieve data for \changed{several tweets}, possibly because they had indeed been blocked by Twitter for exhibiting nefarious activity.
\changed{While several works~\cite{memon2020characterizing, yang2020prevalence, ferrara2020types} have studied the prevalence of COVID-19 misinformation on Twitter and characterized the role of bots in spreading misinformation finding that the fraction of bots sharing misinformation was much higher than those sharing factual information, future work should nonetheless continue to investigate how automatic adversaries spread misinformation in light of the caveat that Twitter might be removing high-engagement tweets before the research community is able to analyze them.}

\changed{Additionally, we were unable to investigate demographic attributes and their impact on engagement with falsehoods. 
This is undoubtedly a dimension that may prove fruitful; Guess et al.~\cite{guess2019less}, for example, investigated specific demographic attributes and their impact on the dissemination of fake news via a survey of individuals and their sharing activity on Facebook. 
They observed that those who consider themselves ``very conservative'' were found to have shared the most articles from domains identified as disseminating false information. 
Additionally, individuals over the age of 65 were observed to share more than twice as many fake news articles than those in the second-oldest age group, suggesting that age is a significant factor in predicting the spread of misinformation. 
}

\changed{Two similar features were deemed negatively correlated with engagement for both misinformation and factual COVID-related tweets: the tweet's length, as measured by word count, and the use of words $>6$ words}.
Historically, we have seen the use of short texts, lots of images, a touch of sex, and a tendency towards sensationalism used as a recipe for propaganda success, leveraged by the KGB, Stasi, and CIA~\cite{rid2020active}. 
The presence of an image and the amount of text (and therefore information that a user must process) in a tweet might be leveraged by both disinformation campaigns and reputable sources alike to help users quickly digest information. 
Additionally, this could denote that users are more likely to engage with an image over words, especially considering that sociolinguistic and sentiment features were not of utmost importance in predicting engagement.
\changed{While we did not measure for the presence of an image, few studies (e.g., \cite{brennen2020beyond, serrano2020nlp}) have conducted exploratory research on visual misinformation videos, and we advise future work to consider this dimension in their work.}

Another limitation of this study is that we \changed{relied on pairwise correlation analysis} to identify the most relevant features \changed{pertaining to misinformation and factual tweets relative to engagement.} 
Our methodology has limitations as it is based on Pearson's correlation coefficient, which measures only linear relationships between the features with each other and with the class variable. 
\changed{
Though we attempted to mitigate this limitation by transforming the variables and calculating the Maximal Correlation, it was nonetheless based on Pearson's correlation.
Furthermore, such as with our decision to use the Mann-Whitney U test to compare the differences between mean ranks of the groups (given that the data and therefore its means were highly skewed and nonparametric), it would have appropriate to leverage Spearman's correlation coefficient, which measures the \textit{rank} correlation.
We also relied only on pairwise comparisons, which are likely far too simplistic in considering the complexities and nuances of large data.
Though this was a driving factor towards considering only strong ($r_{MC,z}\geq 0.50$) correlations, there were many weak to moderate correlations ($0.1 \leq r_{MC,z} < 0.50$) that were overlooked. 
}

\changed{Instead, future work will entail multivariate analyses.}
For example, principal component analysis (PCA) could be used to decorrelate the features before training learning models, as well as to present how many features are needed to explain the dataset, instead of looking at pairwise features (i.e., looking at predictive power that all features in combination are able to provide instead of focusing on single feature predictive value). 
Future work might also seek to understand the feature selection process by checking whether there is a positive or negative correlation between the groups of features (e.g., tweet metadata features and sociolinguistic features). Future studies testing different feature selection methods from both filter (selecting relevant features regardless of a particular classifier) and wrapper (searching for optimal features tailored to a particular model) approaches are warranted.

\section{Conclusion}\label{sec:conclusion}
This paper \changed{curated a dataset of 2.1M COVID-19- and non-COVID-related} misinformation and factual tweets to investigate misinformation as a function of \changed{veracity, content, and engagement}. 
Via the use of statistical and \changed{correlation} analyses, we offer the following conclusions: 
(i) misinformation tweets were less engaging than factual tweets; 
(ii) \changed{features for general and COVID-related tweets varied in correlation to engagement based on veracity; for example, user metadata features (e.g., followers count) were most strong associated with engagement for general misinformation, which COVID-related misinformation correlated most with grammar-related features present in the tweet's text}.
We propose several directions and suggestions for future works on misinformation in the online sphere.
In particular, our insights on what features \changed{can aid with predicting} high engagement can be leveraged for defense approaches against \changed{misinformation}, such as increasing the engagement of factual tweets, especially those coming from verified government accounts and reputable organizations (e.g., WHO, NIH), thus contributing to factual public health information reaching the masses.




\bibliography{bib/main}
\bibliographystyle{ACM-Reference-Format}



\end{document}